\documentclass[sigconf, nonacm]{acmart}

\newcommand\mysys{LazyVLM}

\usepackage{pifont}

\begin{document}

\title{LazyVLM: Neuro-Symbolic Approach to Video Analytics}

\author{Xiangru Jian}
\authornote{Equal contribution to this research.}
\affiliation{%
  \institution{University of Waterloo}
  %\country{Canada}
}
\email{xiangru.jian@uwaterloo.ca}

\author{Wei Pang}
\authornotemark[1]
\affiliation{%
  \institution{University of Waterloo}
  %\country{Canada}
}
\email{w3pang@uwaterloo.ca}

\author{Zhengyuan Dong}
\authornotemark[1]
\affiliation{%
  \institution{University of Waterloo}
  %\country{Canada}
}
\email{zhengyuan.dong@uwaterloo.ca}

\author{Chao Zhang}
\authornotemark[1]
\affiliation{%
 \institution{University of Waterloo}
 %\country{Canada}
}
\email{chao.zhang@uwaterloo.ca}

\author{M. Tamer {\"O}zsu}
\affiliation{%
  \institution{University of Waterloo}
  %\country{Canada}
  }
\email{tamer.ozsu@uwaterloo.ca}

\begin{abstract}

Current video analytics approaches face a fundamental trade-off between flexibility and efficiency. End-to-end Vision Language Models (VLMs) often struggle with long-context processing and incur high computational costs, while neural-symbolic methods depend heavily on manual labeling and rigid rule design. 
In this paper, we introduce  \mysys, a neuro-symbolic video analytics system that provides a user-friendly query interface similar to VLMs, while addressing their scalability limitation.
\mysys\ enables users to effortlessly drop in video data and specify complex multi-frame video queries using a semi-structured text interface for video analytics. 
To address the scalability limitations of VLMs, \mysys\ decomposes multi-frame video queries into fine-grained operations and offloads the bulk of the processing to efficient relational query execution and vector similarity search.
We demonstrate that \mysys\ provides a robust, efficient, and user-friendly solution for querying open-domain video data at scale.
\end{abstract}

\maketitle

\section{Introduction}

With the massive volume of video data in real-world applications, analyzing video content has become increasingly crucial across various domains   \cite{2023-eva-sigmod,equivocal2023demo,sketchql2024wu,2019-blazeit-vldb, romero2023zeldavideoanalyticsusing,2023-vocal-vldb, VOCALExplore2024daum}. A common task in video analytics involves identifying a specific short video segment within a longer video. For instance, consider a surveillance video capturing road traffic: retrieving a moment where a motorcycle is positioned to the right of a bus and then moves to the left of the bus within a few seconds can be instrumental in detecting anomalous behavior in road transportation. 
We refer to this type of query that retrieves multi-frame events as a \textit{video moment retrieval query} (VMRQ).

One category of existing systems for VMRQ processing relies on manual-intensive configurations and interactions. In such systems,  users must define task-specific machine learning models for video content processing, which requires prior domain knowledge. These systems typically employ either a SQL-like query language   \cite{2019-blazeit-vldb,2023-eva-sigmod} or a query-by-example interface   \cite{equivocal2023demo,sketchql2024wu,VOCALExplore2024daum}. The former struggles to express complex video moments due to the need for multiple join expressions or recursive joins to define video moments, while the latter demands extensive user interaction with the video content to perform the search manually (labeling intermediate query results). These designs result in low human efficiency.

\begin{figure*}
    \centering
    \includegraphics[width=\linewidth]{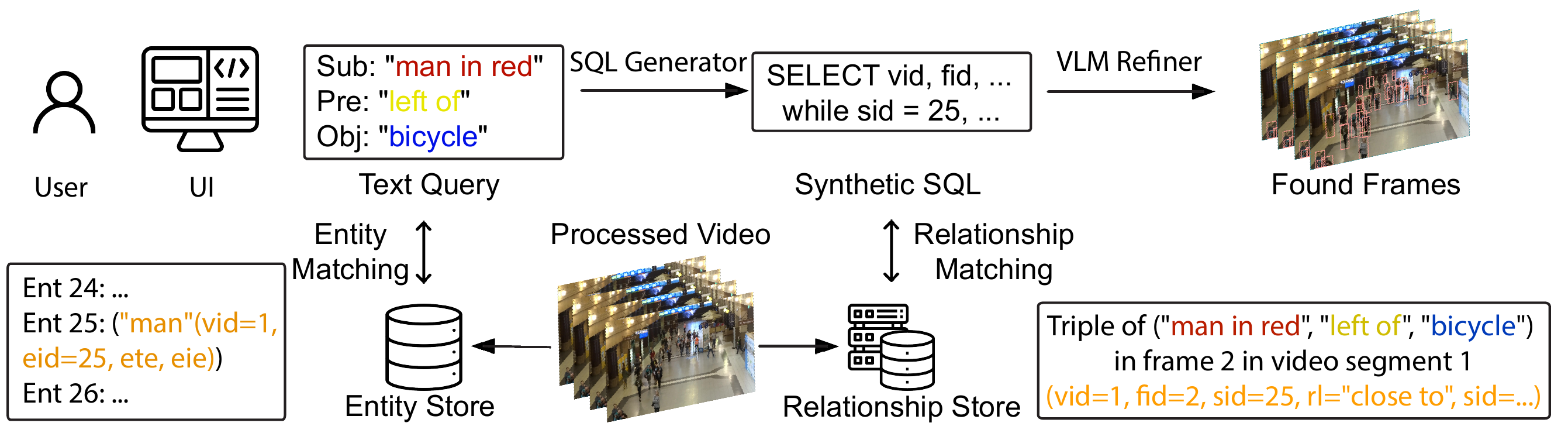}
    \caption{Overview of query processing in \mysys. 
    The diagram illustrates the processing of a semi-structured text query, which includes entity descriptions (\textit{e.g.}, \textit{"man in red"}) and relationship terms (\textit{e.g.}, \textit{"near"}).
    The query is processed through a sequence of stages: entity matching via vector similarity search, SQL query processing to retrieve candidate relationships, relationship verification using a VLM to refine results, and temporal matching to identify the final set of video segments.}
    \label{fig:overview}
\end{figure*}

Alternatively, vision large multimodal models (VLMs)  can be used for video query processing \cite{bai2025qwen25vltechnicalreport}. VLMs offer an intuitive, end-to-end video processing interface: users can simply load a video and input natural language queries into the context window, providing high human efficiency. However, VLMs fall short when processing video queries at scale due to several limitations.
First,  VLMs' inference  is highly time-consuming due to the underlying self-attention mechanism, which requires $O(n^2)$ time complexity where $n$ is the total number of tokens in the context window—this overhead is primarily dominated by video length.
Second, the autoregressive nature of decoder-only VLMs enforces a serial computation mechanism, significantly limiting opportunities for parallel processing.
Finally, executing ad-hoc exploratory queries or video updates, such as adding new videos, necessitates repeated processing of the entire context window, further exacerbating inefficiencies. Therefore, using VLMs out of the box for video query processing leads to low system efficiency.

To overcome the above issues, we propose \mysys, a neuro-symbolic approach designed for scalable video analysis.
\mysys\ introduces a semi-structured text interface for defining video events, allowing users to describe visual content in frames using subject-predicate-object (SPO) triples, where each element (S, P, or O) is specified in text (see Example \ref{ex:query}).
Video can be loaded into \mysys\ without defining task-specific models, \textit{i.e.}, by simply dropping in the video with no additional effort required.
By offering an interface similar to that of VLMs, \mysys\ improves human efficiency, compared to traditional manual-intensive systems.

\mysys\ significantly reduces the computation cost, compared to out-of-box VLMs, thereby improving system efficiency for video analytics. 
This is achieved by generating structured views of video data—specifically, scene graphs for frames—and embedding nodes (representing entities) within these graphs.
\mysys\ then decomposes video queries into semantic search and symbolic search.
Semantic search is based on embedding vectors and is dedicated to searching an entity (either a subject or an object in the SPO triples).
Symbolic search is based on relational queries and focuses on verifying the existence of relationships (predicates between subjects and objects in SPO triples).
Additionally, \mysys\ leverages VLMs to refine the results of relational queries. However, since most computation is offloaded to semantic search and vector search, VLMs are used in a lightweight manner. Instead of processing an entire video, \mysys\ selectively applies VLMs to individual frames that have been verified through relational queries—dramatically reducing computational overhead.
Thanks to its fine-grained query decomposition, \mysys\ enables parallel computing, allowing multiple relational queries, vector searches, and VLM refinements to be executed simultaneously.
Moreover, structured views and embedding vectors are precomputed once and stored. This makes \mysys\ update-friendly, supporting incremental updates by inserting new structured views and vectors, eliminating the need to reprocess the entire video from scratch.

\mysys\ enhances both human efficiency and system efficiency by combining the strengths of symbolic search, semantic search and VLMs. This integration enables a powerful and efficient framework for querying open-domain video data. The remainder of this paper provides a detailed overview of \mysys's architecture and interaction mechanisms.

\section{System Overview}

Figure~\ref{fig:overview} illustrates the processing pipeline of \mysys.
Video data is first preprocessed to generate structured views, which are stored in the \textbf{Entity Store} and \textbf{Relationship Store} (see Section~\ref{subsec:video_proc} for details).
\mysys\ provides a semi-structured text interface that allows users to query the loaded video data.
At query time, \mysys\ employs a series of components to retrieve relevant information from the two stores, followed by the use of a VLM to refine the pruned set of candidate results.
The remainder of this section describes the key components of \mysys\ in detail.

\subsection{Query Interface and Functionality}
The query interface of \mysys\ is based using SPO triples to describe the visual content in video frames. We use the following example to illustrate how users can describe multi-frame events in a natural and intuitive manner.

\begin{example}\label{ex:query}
Consider the following query that defines a complex event: a man with a backpack is near a bicycle, and another man in red clothes moves from the left of the bicycle to the right of the bicycle after more than 2 seconds. This query is formally specified in the following steps in \mysys.
\textbf{(1) Entity Description:} $E=\{e_1, e_2, e_3\}$, with $e_1.\text{text} = $\textit{"man with backpack"}, $e_2.\text{text} = \text{\textit{"bicycle"}}$, and $e_3.\text{text} = \text{\textit{"man in red"}}$. These defined entities can serve as either a subject or an object in SPO triples.
\textbf{(2) Relationship Description:} $R=\{r_1, r_2, r_3\}$, where $r_1.\text{text} = \text{\textit{"is near"}}$, $r_2.\text{text} = \text{"\textit{leftOf}"}$, and $r_3.\text{text} = \text{"\textit{rightOf}"}$. These relationships function as predicates in SPO triples.
\textbf{(3) Frame Description:} $F = (f_0, f_1)$, where $f_0 = \{(e_1, r_1, e_2), (e_3, r_2, e_2)\}$ represents \textit{"man with backpack is near bicycle; man in red is on the left of bicycle"}, and $f_1 = \{(e_1, r_1, e_2), (e_3, r_3, e_2)\}$ represents \textit{"man with backpack is near bicycle; man in red is on the right of bicycle"}.
\textbf{(4) Temporal Constraint:} $f_1 - f_0 > 4$, ensuring that the second frame occurs more than 2 seconds after the first, assuming a frame rate of 2 frames per second.
\end{example}

In \mysys, input video is automatically divided into non-overlapping video clips, \textit{i.e.}, short segments lasting a few seconds or minutes, with the length defined by the user. \mysys\ then processes an input query, \textit{e.g.}, the one in Example \ref{ex:query}, by searching for events specified in the query, retrieving, and returning the clips that contain the detected event.

\textit{Functionality Supported.} \mysys\ supports a comprehensive range of core video analytics functionalities.
At the object level, \mysys\ supports (1) \textit{object detection}, identifying and localizing entities such as cars, people, and bicycles within video frames, and (2) \textit{object tracking}, which follows these objects across multiple frames, facilitating trajectory-based analysis. 
In addition, \mysys\ includes (3) \textit{attribute recognition}, allowing queries based on object properties like color or size.
At the relationship level, \mysys\ includes (4) \textit{relationship detection}, which captures spatial and interaction-based relationships (\textit{e.g.}, car near pedestrian or person holding an object).
Beyond basic object and relationship analytics, \mysys\ supports advanced query operations: (5) \textit{conjunction queries} allow users to specify multiple conditions (\textit{e.g.}, detecting a person and a vehicle in the same frame), while (6) \textit{sequencing queries} enforce temporal order between events (\textit{e.g.}, detecting a person walking before entering a car). 
Additionally, (7) \textit{window queries} constrain events within a defined time duration (\textit{e.g.}, detecting a car stopping within 10 seconds after a pedestrian appearing).

\textit{User Interaction Flow.}
User interaction is straightforward with \mysys.
\textbf{(1) Upload Video Dataset:} Users begin by uploading their videos. \mysys\ automatically segments the videos, extracts visual features, and builds the corresponding structured views.
\textbf{(2) Compose a Query:} 
    Through the intuitive text interface, users describe the event of interest. For example, to retrieve clips where \textit{“a man with a backpack is near a bicycle, and a man in red moves from being on the left of the bicycle to the right of the bicycle after more than 2 seconds”}, users enter the corresponding entity descriptions, relationship descriptions, frame information, and temporal constraints, as shown in Example \ref{ex:query}. 
    The query is then submitted to the system, and users can view the results after the query execution in the interface. Detailed steps are presented in Section~\ref{sec:demo}.

\subsection{Video Preprocessing} \label{subsec:video_proc}
Video preprocessing converts raw video content into a structured, searchable format through the following integrated stages.

\textit{Video Segmentation.}
When a video dataset is uploaded, each video $V$ is automatically partitioned into a sequence of non-overlapping segments, $V = (v_1, v_2, \dots, v_n)$. Each segment $v_i$ comprises a fixed number of frames, $v_i = (f_1, f_2, \dots, f_m)$, facilitating parallel processing and management of long videos.

\textit{Content Extraction.}
Each video segment is processed to extract detailed visual information. To represent the visual content of each frame, we construct a scene graph that encodes the frame as a set of subject–predicate–object (SPO) triples in the form (subject, predicate, object). For example, a frame may include triples such as (\textit{"man with backpack", "near", "bicycle"}) and (\textit{"bicycle", "on", "sidewalk"}). We employ the IETrans~\cite{ietrans2022zhang} model to generate these scene graphs. 
For extracted triples, each subject or object—internally referred to as an entity—is represented as a tuple $(vid, eid, ete, eie)$, and all entities are stored in the \textbf{Entity Store}.
$vid$ denotes the identifier of the video segment containing the entity; $eid$ is a unique entity identifier within the segment, obtained through entity tracking using YOLOv8~\cite{yolov8_ultralytics}; $ete$ is a text embedding derived from the entity’s description using the e5-mistral-7b~\cite{wang2024improvingtextembeddingslarge} model; and $eie$ is an image embedding that captures the entity’s visual appearance, generated by the VLM2Vec~\cite{jiang2025vlmvec} model. 
In addition to entity representations, inter-object relationships are captured as tuples $(vid, fid, sid, rl, oid)$, which are stored in the \textbf{Relationship Store}. 
$vid$ and $fid$ refer to the video segment and frame identifiers, respectively; $sid$ and $oid$ are the unique entity identifiers of the subject and object involved in the relationship; and $rl$ denotes the relationship label, \textit{e.g.}, \textit{"near"}.

\subsection{Query Processing}
Users express an event query as a four-part specification, including entity descriptions, relationship descriptions, frame-level specifications, and temporal constraints, as illustrated in Example~\ref{ex:query}.
The query engine of \mysys\ processes the query through a series of stages: \textit{Entity Matching}, \textit{SQL Query Generation}, \textit{Relationship Matching and Refinement}, and \textit{Temporal Matching}.

\textit{Entity Matching.}
For each entity defined in the query, a vector similarity search is performed to match the textual description of the entity against the embeddings stored in the Entity Store. The result is a set of candidate entities for each query entity, represented as $(vid, eid)$ pairs, \textit{i.e.}, matched entities in specific video segments.

\textit{SQL Query Generation.}
Based on the entity matching results, the query engine automatically generates SQL queries to retrieve candidate frames from the Relationship Store. These queries filter rows by matching entity identifiers, \textit{i.e.}, $(vid, eid)$ pairs, and they return frames that potentially include the specified entities. The output is a set of rows from the Relationship Store corresponding to each query entity, denoted as candidate frames.

\textit{Relationship Matching and Refinement.}  
For each SPO triple in the query, the query engine performs a join between the candidate frames for the subject and object, based on shared $vid$ and $fid$ values. This identifies potential relationships between subject and object candidates. Following this coarse-grained match, a refinement step is applied to further verify relationships. Specifically, a lightweight local VLM (\textit{e.g.}, Qwen-2.5-VL 7B~\cite{bai2025qwen25vltechnicalreport}) is used for the verification. Optionally, users may apply cost-efficient closed-source VLMs (\textit{e.g.}, GPT-4o-mini) for deployment-friendly setups. This refinement ensures that the relationships specified in the query are visually and semantically grounded in the candidate frames.
Finally, the engine joins the refined relationship results to identify frames where all specified SPO triples in a query frame co-occur, producing a set of candidate video frames for each query frame.

\textit{Temporal Matching.}
In the final stage, the engine checks whether the candidate video frames satisfy the temporal constraints defined in the query. This involves join over frame identifiers to ensure compliance with temporal logic, \textit{e.g.}, $f_1 - f_0 > 4$. Ultimately, video identifiers including candidate video frames that satisfy the temporal conditions are returned as the final query results.

A major advantage of the proposed query processing pipeline is that each step is inherently parallelizable. For instance, entity matching tasks can be executed in parallel, as they are independent of one another. 
In addition, the query engine applies VLMs only for fine-grained relationship verification on a pruned set of candidate frames, rather than processing all video frames, making it significantly more lightweight compared to end-to-end VLM-based approaches for video query processing. 

\section{Demonstration} \label{sec:demo}

\begin{figure*}[htbp]
    \centering
    \includegraphics[width=\linewidth]{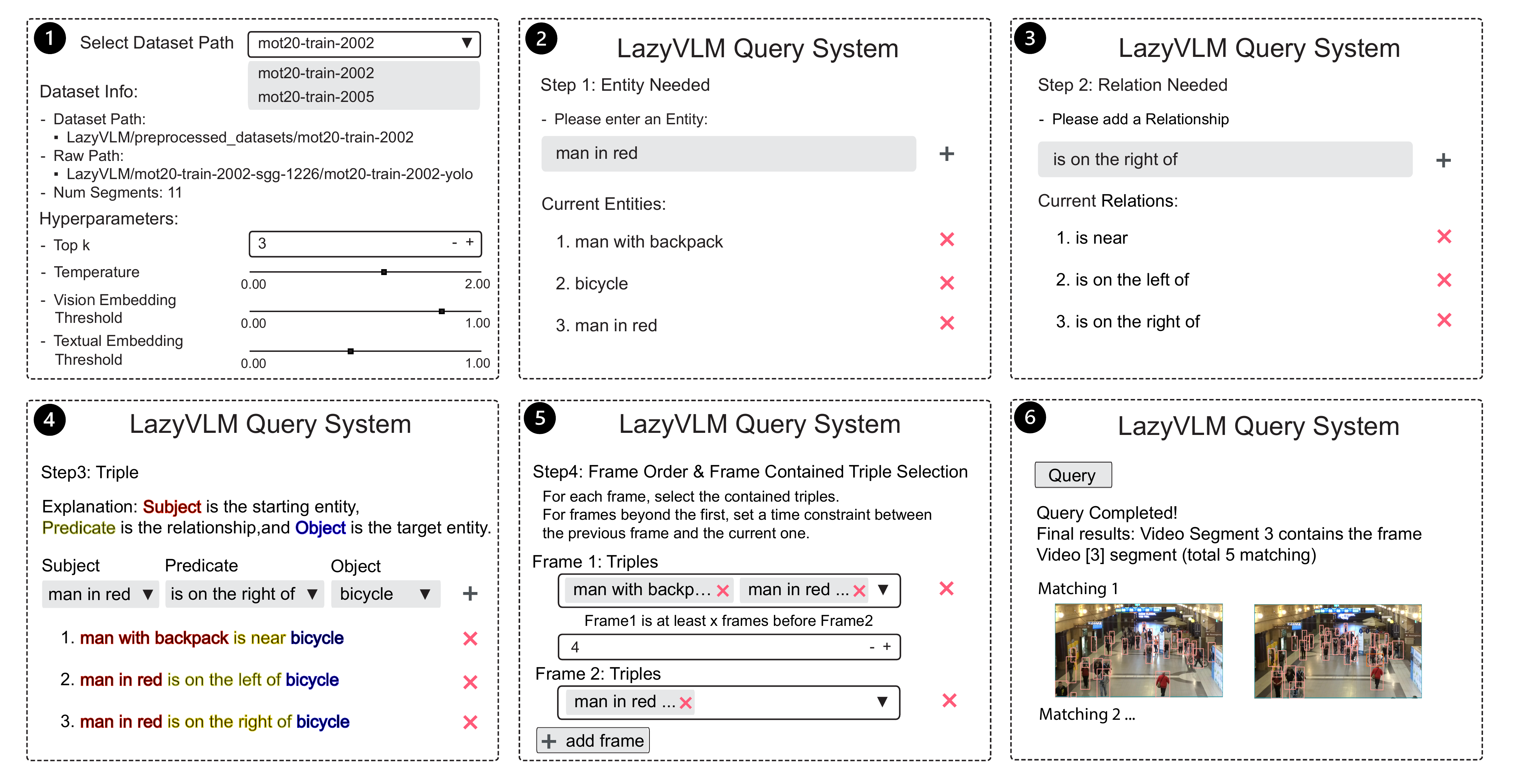}
    \caption{Pipeline of user interactions in \mysys\ for specifying and executing a video query: 
     \textbf{Step \ding{182}: Load Dataset and Enter Hyperparameters};
    \textbf{Step \ding{183}: Enter Entities};
    \textbf{Step \ding{184}: Enter Relationships};
    \textbf{Step \ding{185}: Enter Triples};
    \textbf{Step \ding{186}: Enter Frames and Temporal Constraints}; and
    \textbf{Step \ding{187}: Query Execution and Presentation of Results}.}
    \label{fig:pipeline_interface}
\end{figure*}

We demonstrate \mysys\ using the MOT20~\cite{MOTChallenge20} and TAO~\cite{Dave:2020:ECCV} datasets. Specifically, we use Example~\ref{ex:query} to perform a query on the MOT20-02 dataset. 
We guide the users through the following detailed steps, illustrating the interactive and intuitive interface of \mysys.

\noindent \textbf{Step \ding{182}: Load Dataset and Enter Hyperparameters.} Users select the MOT20-02 dataset path from a dropdown menu. The interface shows dataset metadata, including the total number of video segments (\textit{e.g.}, 11 segments for MOT20-02), as well as paths for preprocessed and raw data. Users then configure several hyperparameters for query execution: the \textit{top-k} parameter \textit{(e.g.}, top 3 results), the \textit{temperature} for controlling search strictness, and the thresholds for \textit{vision embedding} and \textit{textual embedding} similarity searches, which affect entity and relationship matching accuracy.

\noindent \textbf{Step \ding{183}: Enter Entities.} Users input descriptive text labels for the entities involved in their query via a dedicated input field. In the provided example query, users define entities like \textit{"man with backpack," "bicycle,"} and \textit{"man in red"}. These entities are then listed on the interface and can be reviewed or removed if necessary.

\noindent \textbf{Step \ding{184}: Enter Relationships.} Users specify relationships that describe interactions or spatial positions between entities, such as \textit{"is near," "is on the left of,"} or \textit{"is talking to"}. Each entered relationship appears on the interface for adjusting before proceeding.

\noindent \textbf{Step \ding{185}: Enter Triples.} Users construct SPO triples to precisely define the interactions between entities. The interface supports triple formation by allowing users to select from previously entered entities and relationships. For instance, users define triples like \textit{"man with backpack is near bicycle," "man in red is on the left of bicycle,"} and \textit{"man in red is on the right of bicycle"}. Each formed triple is displayed for user verification.

\noindent \textbf{Step \ding{186}: Enter Frames and Temporal Constraints.} Users organize the defined triples into specific frames according to their query's temporal constraint. For each frame, users explicitly select which triples it contains from a dropdown list. For the example provided, Frame 1 is set to contain triples \textit{"man with backpack is near bicycle"} and \textit{"man in red is on the left of bicycle,"} while Frame 2 includes \textit{"man with backpack is near bicycle"} and \textit{"man in red is on the right of bicycle."} Users also define temporal constraints, such as specifying Frame 1 to occur at least 4 frames before Frame 2 since we have 2 frames per second in the video.

\noindent \textbf{Step \ding{187}: Query Execution and Presentation of Results.} Upon completing the query setup, users initiate query execution by clicking the \textit{"Query"} button. \mysys\ processes the query and displays matching results. The results detail the precise video segments and exact frame identifiers corresponding to each user-defined frame. 

\section{Conclusion}
We present \mysys, a neuro-symbolic video analytics system that efficiently supports complex, multi-frame queries through intuitive, text-based interaction. Our demonstration highlights \mysys’s scalability, accuracy, and usability in practical, open-domain video analytics scenarios.

\bibliographystyle{ACM-Reference-Format}

\bibliography{bibliography}
\end{document}